\documentclass[aps,prb,twocolumn]{revtex4-1}
\usepackage[english]{babel}
\usepackage[percent]{overpic}

\usepackage{verbatim}
\usepackage[utf8]{inputenc}
\usepackage{amsmath,amssymb,amsfonts}
\usepackage{color}
\usepackage{hyperref}
\usepackage{filecontents}
\usepackage{graphicx}
\usepackage{soul}
\usepackage{bm}
\begin{document}     
\title{Polarized heat current generated by quantum pumping in two-dimensional topological 
insulators}
\author{F. Ronetti$^{1,2,3}$, M. Carrega$^{4}$, D. Ferraro$^3$, J. Rech$^3$, T. Jonckheere$^3$, T. Martin$^3$, M. Sassetti $^{1,2}$}
\affiliation{ $^{1} $ Dipartimento di Fisica, Universit\`a di Genova, Via Dodecaneso 33, 16146, Genova, Italy.\\
$^2$ CNR-SPIN, Via Dodecaneso 33, 16146, Genova, Italy.\\
$^3$ Aix Marseille Univ, Université de Toulon, CNRS, CPT, Marseille, France\\
$^4$ NEST, Istituto Nanoscienze-CNR and Scuola Normale Superiore, I-56127 Pisa, Italy.
} 

\begin{abstract}
	We consider transport properties of a two dimensional topological insulator in a double quantum point contact geometry in presence of a time-dependent external field. In the proposed setup an external gate is placed above a single constriction and it couples only with electrons belonging to the top edge. This asymmetric configuration and the presence of an ac signal allow for a quantum pumping mechanism, which, in turn, can generate finite heat and charge current in an unbiased device configuration. A microscopic model for the coupling with the external time-dependent gate potential is developed and the induced finite heat and charge current are investigated.
We demonstrate that in the non-interacting case, heat flow is associated with a single spin component, due to the helical nature of the edge states, and therefore a finite and polarized heat current is obtained in this configuration. The presence of e-e interchannel interactions strongly affects the current signal, lowering the degree of polarization of the system. Finally, we also show that separate heat and charge flows can be achieved, varying the amplitude of the external gate.
\end{abstract}

\maketitle

\section{Introduction}
The fast development in nanotechnologies has spurred the investigation of quantum effects in electronic devices at submicron scale~\cite{dassarma97, vignale04, giamarchi03, martin05, pekola15}. In past years, the attention has been devoted mostly to the study of charge transport and related phenomena in electronic nanocircuits~\cite{buttiker93, bocquillon14, feve07, giazotto11, carrega11,wahl14}. Here spectacular effects have been predicted and also experimentally observed, including quantum interference pattern~\cite{dassarma97}, single electron injection and control in quantum conductors~\cite{bocquillon14, feve07, kamata14,ferraro15} and coherent behavior in hybrid devices~\cite{martin05, pekola15, giazotto11, giazottorev, padurariu15, ferraro15b}.\\
Conversely, the question how thermodynamic aspects such as heat transport, power conversion and energy exchange work at the nanoscale has posed new interesting challenges~\cite{esposito09, campisi11, giazottorev, lirev, casatirev,sancheznjp,sanchezprl,thierschmann16}. A precise control and manipulation of heat flows in quantum conductors can lead to interesting applications, i.e., new logic devices based purely on thermal transport. Towards this goal, pioneering experiments have achieved the phase control of coherent  thermal transport in hybrid quantum systems~\cite{pekola15, giazottorev, jezouin13, perez14, altimiras12, meschke06}.\\
In this respect, the possibility to tune and control some physical parameters by means of external fields and time-dependent potential is of great interest~\cite{ludovico14, ludovico16, carrega16, ludovicorev, ludovicoconf, arrachea11}. Indeed,  periodically driven system can be immediately associated to thermal machines or heat engines which are based on cyclic operations. Different papers focused on the effects induced by an external driving field both in closed and open quantum systems.
For example, the influence of an ac field on the efficiency of thermoelectric setups or heat pumping mechanism on quantum dot based devices have been investigated~\cite{ludovico14, ludovico16, ludovicorev, calzona16, moskalets14, battista14}.\\
Dynamical aspects of time-dependent energy and heat flows have been addressed in order to understand fundamental thermodynamical problems such as energy exchange, work distribution and entropy production (see reviews ~\cite{casatirev, ludovicorev} and references therein).\\
Interestingly, locally applied time-dependent voltages can lead to heat pumping mechanisms in quantum systems~\cite{giazotto11, ludovicorev, Moskalets13, arrachea07, moskalets02, thouless, brouwer}.
The peculiar characteristic of heat pumps is that a direct heat current is generated by a purely ac drive which acts against some present thermal gradients.
Moreover, it can also generate dc finite current signals even in a purely equilibrium situation, in absence of any thermal or voltage bias~\cite{ludovicorev, ren10, quansing10}.\\
Recently,  a lot of attention has also been put on the interplay between heat and spin transport properties, in the emerging research field of spin caloritronics \cite{bauer10, bauer12,kolenda16}. Thermally driven spin sources are very interesting for the improvement of spintronic devices, which are believed to be more efficient with respect to their electronic counterparts~\cite{giazottorev, casatirev, murakami03}.
Striking experimental results in this field have extended thermoelectric  concept to spin transport by the observation of spin-Seebeck \cite{uchida08} and spin-Peltier \cite{flipse12} effects in magnetic systems. Intriguingly, spin-dependent effective temperatures of electrons were observed in nanopillar spin valves powered by a heat source, thus showing that electron heat conduction can depend on the spin degree of freedom \cite{dejene13, heikkila10}.\\
Among all mesoscopic systems for the investigation of spin transport properties, a relevant role is certainly played by two-dimensional topological insulators (2dTIs)\cite{dolcetto15,zhang11,bernevig06,konig07}. These system are constituted by a wide bulk gap and charge and energy transport are mediated by the presence of edge states.
These are constituted by two counter-propagating electronic states with opposite spin polarization (spin-momentum locking), and are also called helical edge states~\cite{dolcetto15, schmidt11}. Moreover, as long as time-reversal symmetry is preserved, electrons flowing along these edge states are topologically protected against backscattering and transport occurs in the so-called ballistic regime~\cite{dolcetto15, zhang11, bernevig06}.
Experimental evidences of 2dTIs were reported in CdTe/HgTe \cite{roth09} and InAs/GaSb \cite{knez11,du15} quantum wells and have been predicted for a large class of new materials which relies on the presence of strong spin-orbit coupling~\cite{reis16, zhang16, zhou15}.
These systems constitute an interesting playground to study coherent heat transport in which the spin degree of freedom can play a non-trivial role~\cite{ludovicorev, ronetti16}.
Moreover, the unavoidable presence of e-e interactions in helical edge states~\cite{dolcetto15, li15, strom09, hou09, dolcetto12} can dramatically affect the dynamics and have important consequences on thermal transport properties~\cite{ronetti16}.\\

In this work we focus exactly on these issues. We consider  a 2dTI device in a double quantum point contact geometry~\cite{dolcini11, sternativo13, ferraro13, chamon97, huang13, vannucci15}. In this setup tunneling of electrons is allowed by the presence of constrictions and can lead to quantum interference effects.
We also introduce a model to  couple electrons of the top edge to a local time-dependent gate potential located above a single QPC. Our aim is to investigate the generation of a direct heat current by the time-dependent drive along the edge of a 2dTI. We demonstrate that the presence of an ac field can generate dc finite current signals in absence of thermal or voltage bias, relying on a quantum pumping mechanism.
We inspect the interplay with the peculiar spin properties of helical edge states, demonstrating that in the proposed setup a polarized heat current can be enstaured, carried solely by a single spin species.
Finally, the presence of e-e interactions can strongly modify the behavior of the heat pumped current in this device. Interestingly, it turns out that by looking at the resulting oscillating patterns one can identify the presence of e-e interactions in the system.\\

The paper is organized as follows. In Sec. \ref{setup}, we present the setup and describe the various contributions to the Hamiltonian of the system. In Sec. \ref{currents}, we define charge and heat currents and evaluate their average values. Sec. \ref{results} is devoted to the discussion of our main results. Finally, in Sec. \ref{conclusions}, we draw our conclusions.

\section{Model and setup \label{setup}}
We consider a 2dTI connected to two reservoirs at equilibrium kept at the same chemical potential $\mu$ and temperature $T$.
\begin{figure}[h]
\centering
\includegraphics[scale=0.35]{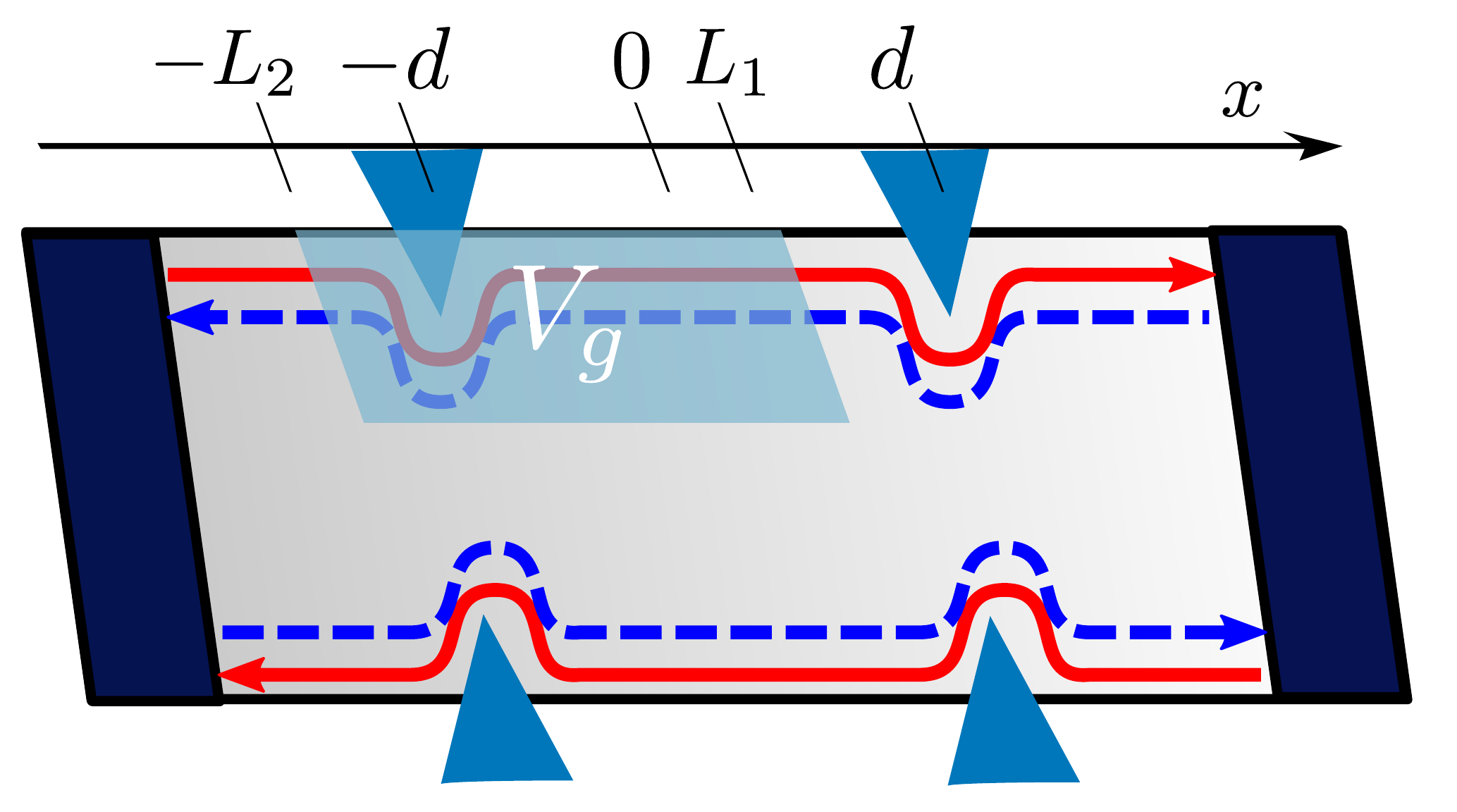}
\caption{(Color online) Scheme of the proposed setup. A 2dTI is connected to two reservoirs with chemical potential $\mu$ and temperature $T$. On each edge, the direction of propagation of spin $\uparrow$ (solid red line) and spin $\downarrow$ (dashed blue line) electrons is opposite. Two quantum point contacts, placed at $x=\pm d$, allow for interedge tunneling. An ac external gate voltage $V_g(x,t)$ is capacitively coupled to the top edge in the region around $x\sim -d$ between $-L_2$ and $L_1$ ($L_1$ can take value between $-d < L_1< d$).}
\label{giunzione1}
\end{figure}
Here, electrons can propagate along the edges and, in the absence of interactions, have well-defined chirality of right ($R$) and left ($L$) moving particles~\cite{dolcetto15, zhang11}. Moreover, the spin-momentum locking property of 2dTI constrains the direction of propagation and the spin projection of electrons: in the top edge the spin of $R$-electrons is $\uparrow$ and the spin of $L$-electrons is $\downarrow$ and viceversa for the bottom edge. Top and bottom edges are separated by macroscopic distances and inter edge tunneling is only possible in presence of some constrictions. Here, we consider a double quantum point contact (QPC) geometry~\cite{ferraro13, chamon97, huang13}, where the two constrictions are placed at $x=\pm d$, as schematically shown in Fig.~\ref{giunzione1}.
Electrons in the top edge are capacitively coupled to an external gate potential $V_g(x,t)$, confined in a region $-L_2<x<L_1$, see Fig.~\ref{giunzione1}. The presence of a time-dependent gate potential, assumed to be periodic in time for simplicity, is crucial in order to generate dc finite current signals, without any external dc bias~\cite{ludovicorev, brouwer}.
The total system can be described by the following Hamiltonian
\begin{equation}
\label{htot}
\hat{H}=\hat{H}_{HLL} + \hat{H}_{tun} + \hat{H}_g,
\end{equation}
where $\hat{H}_{HLL}= \hat{H}_{free} + \hat{H}_{int}$ describes the interacting edge states of the 2dTI.
The free contribution associated to fermionic channels can be written as
\begin{equation}
\hat{H}_{free}= \sum_{r=R/L}\sum_{\sigma=\uparrow,\downarrow} \hat{H}_{r,\sigma},
\end{equation}
with (from now on $\hbar=1$)
\begin{equation}
\label{ham_free_rs}
\hat{H}_{r,\sigma}=-i\xi_r v_{\rm F}\int dx\hspace{0.5mm}e^{-i\xi_r k_{\rm F}x}:\hat{\psi}^{\dagger}_{r,\sigma}(x)\partial_x\left(e^{i\xi_r k_{\rm F} x}\hat{\psi}_{r,\sigma}(x)\right):,
\end{equation}
where $v_{\rm F}$ and $k_{\rm F}=\mu/v_{\rm F}$ are the Fermi velocity and Fermi momentum respectively, $\xi_{R/L}=\pm1$, and $\hat{\psi}^\dagger_{ r,\sigma}(x)$ the corresponding electron creation operator.
In the following, electron interactions on each edge are treated
assuming the breaking of galileian invariance, i.e., in the absence of intra channel interactions \cite{giamarchi03, schmidt11, martin05, miranda03, geissler15}, and the corresponding term reads
\begin{equation}
\hat{H}_{int}=\sum_{\sigma=\uparrow,\downarrow} 2g_2\int dx \hspace{1mm}\hat{\rho}_{ R,\sigma}(x)\hat{\rho} _{L,-\sigma}(x),
\end{equation}
where $g_2$ is the coupling constant describing the interchannel interaction strength (for sake of convenience, we have adopted the same  label as in the Luttinger liquid language\cite{giamarchi03}).
In the above equation the density operator is denoted as $\hat{\rho} _{ r,\sigma}(x)=:\hat{\psi}^{\dagger}_{ r,\sigma}(x)\hat{\psi}_{ r,\sigma}(x):$.
Following standard bosonization prescriptions\cite{giamarchi03, martin05, miranda03} (see Appendix \ref{zero} for more details), $\hat{H}_{HLL}$ can be written in a diagonal form in terms of chiral bosonic field operators $\hat{\phi}_{\pm}^{(\ell)}(x)$ as 
\begin{equation}
\label{hamdiag}
\hat{H}_{HLL}=\sum_{\ell=1,2} u\int dx \hspace{1mm}\left[\left(\partial_x\hat{\phi}_{+}^{(\ell)}(x)\right)^2+\left(\partial_x\hat{\phi}_{-}^{(\ell)}(x)\right)^2\right]
,
\end{equation}
where  $u = v_{\rm F} \frac{2K}{1+K^2}$ represents the renormalized velocity, expressed in terms of the dimensionless interaction parameter $K=\sqrt{\frac{2\pi v_{\rm F}-g_2}{2\pi v_{\rm F}+g_2}}$\cite{geissler15} and $\ell=1,2$ identifies the top/bottom edge respectively. It is worth to note that in this work we restrict the analysis to the range of interaction strengths given by $1/\sqrt{3} \leq K \leq 1$, for which it is possible to show that single electron tunneling is the dominant process (in the renormalization group sense), both in the single and double QPC geometry\cite{schmidt11, chamon97, huang13, ronetti16}.\\
Tunneling events between the two edges are allowed in the double QPC geometry under consideration. For sake of simplicity, here we focus on spin-preserving tunneling events described by \cite{dolcini11, sternativo13,dolcetto12,ferraro13}
\begin{equation}
\label{SP_ham}
\hat{H}_{tun}=\sum\limits_{\sigma=\uparrow,\downarrow}\int\hspace{1mm}dx \hspace{1mm}h(x) e^{i2k_{\rm F} x}\hat{\psi}^{\dagger}_{ L,\sigma}(x)\hat{\psi}_{R,\sigma}(x)+ \text{H}.\text{c}.,
\end{equation}
with $h(x)=\Lambda\sum\limits_{p=\pm 1}\delta(x-pd)$ the space dependent amplitude describing the double constrictions depicted in Fig.\ref{giunzione1} (here $\Lambda$ represents the constant tunneling amplitude).An analogous analysis can be carried out considering spin-flipping tunneling at the QPCs: however, we expect this contribution to be smaller with respect to the spin-preserving one\cite{dolcini11,ojanen11}.\\
Finally, we assume that the external gate potential $V_g(x,t)$ is capacitively coupled to the electron density of the top edge via
\begin{equation}
\label{Ham_gate_giunzione1}
\hat{H}_g=-e\int\limits dx\hspace{1mm} V_g(x,t)\left(\hat{\rho}_{R,\uparrow}(x)+\hat{\rho}_{L,\downarrow}(x)\right),
\end{equation}
with the time-dependent potential
\begin{equation}
\label{ham_gate}
V_g(x,t)= \left[\theta\left(x+L_2\right)-\theta\left(x-L_1\right)\right] V(t).
\end{equation}
Here, $L_1,L_2$ are linked to the dimension of the gate, see Fig.\ref{giunzione1}, and $V(t)=V(t+{\cal T})$ is a periodic time-dependent drive, with period $\mathcal{T}$.
In particular, we will consider the gate located around $x \sim -d$, i.e., $-L_2 < - d$ ($L_2>0$) and $-d < L_1 <d$, as shown in Fig.~\ref{giunzione1}.
Finally we note that the chosen configuration, with the external  gate located around a single QPC at $x\sim -d$, naturally guarantees the required asymmetry needed for a quantum pumping mechanism~\cite{giazotto11, thouless, brouwer}.
\section{Average pumped currents \label{currents}}
We are interested in the study of transport properties in the setup introduced above. We recall that the 2dTI is connected to two reservoirs, which we assume to be at equilibrium at the same chemical potential $\mu$ and same temperature $T$. Nevertheless finite dc current signals can be generated due to the presence of the time-dependent gate potential. Indeed, as already mentioned, the ac field supplied by the external gate above one single QPC can lead to a quantum pumping mechanism, which, in turn, generates dc finite currents.
Here, we investigate this aspect, by studying the pumped currents which flow through the 2dTI.
In particular we will consider the generation of heat current $I_q$ and charge current $I_c$, focusing on their dc components, extracted from the finite temperature current contributions by further averaging the signals over one period of the gate potential $V(t)$.\\
It is possible to show, see Appendix~\ref{zero}, that in the dc limit the only possible contributions to pumped currents are due to backscattering. The latter are present whenever the edges are coupled and tunneling events occur, i.e., in the proximity of the two QPCs.
Therefore, in this section we focus on the evaluation of the backscattering contribution to the average currents (we thus introduce a related index $BS$).
Backscattering charge current $\hat{I}_c^{BS}$ can be defined as
\begin{equation}
\label{charge_N}
\hat{I}_c^{BS}\equiv-e\left(\hat{I}^{BS}_{N\uparrow}+\hat{I}^{BS}_{N\downarrow}\right),
\end{equation}
    where $(-e)$ is the electron charge ($e>0$) and we have identified the particle current per spin component, which is given by
    \begin{equation}
\label{spin_curr_num}
\hat{I}^{BS}_{N\sigma}=\frac{i}{2}\left[\hat{H}_{tun},\hat{N}_{R,\sigma}-\hat{N}_{L,\sigma}\right],
\end{equation}
with the particle number operator $\hat{N}_{r,\sigma}=\int dx~\hat{\rho}_{r,\sigma}(x)$.
Using Eq.~(\ref{SP_ham}) we can thus write
\begin{equation}
\label{p_op_ferm}
\hat{I}^{BS}_{N\sigma}=i\int dx\hspace{1mm} h(x) e^{i 2 k_{\rm F} x}\hat{\psi}_{L,\sigma}^{\dagger}(x)\hat{\psi}_{R,\sigma}(x)+\text{H}.\text{c}.
\end{equation} 
Analogously, the backscattering heat current flowing between the two reservoirs can be written as the sum of the heat current per spin direction  
\begin{equation}
\label{heatback1}
\hat{I}_q^{BS}\equiv \hat{I}_{q\uparrow}^{BS}+\hat{I}_{q\downarrow}^{BS},
\end{equation}
with
\begin{equation}
\label{heatback}
\hat{I}_{q\sigma}^{BS}\equiv \frac{i}{2}\left[\hat{H}_{tun}, \hat{Q}_{R,\sigma}-\hat{Q}_{L,\sigma}\right],
\end{equation}
where $\hat{Q}_{r,\sigma}=\hat{H}_{r,\sigma}-\mu\hat{N}_{r,\sigma}$.
The expression in Eq. \eqref{heatback}, related to the heat current contribution per spin component, in terms of fermionic fields is
\begin{equation}
\label{h_op_ferm}
\hat{I}_{q\sigma}^{BS}=\frac{v_{\rm F}}{2}\int dx \hspace{1mm}h(x)e^{i 2 k_{\rm F} x}\partial_x\left(\hat{\psi}^{\dagger}_{L,\sigma}(x)\hat{\psi}_{R,\sigma}(x)\right).
\end{equation}
It is worth to underline that Eq. \eqref{heatback1} corresponds to the net amount of heat flow exchanged between $R$-channel and $L$-channel due to backscattering processes and it does not give any information on the heat which locally enters into each reservoir separately.\\
Using standard perturbative approach~\cite{martin05, ferraro13, ronetti16} in the tunneling Hamiltonian, the average pumped currents can be calculated.
At the lowest order in the tunneling one has the finite temperature currents (here $\nu=n,q$ for particle-number and heat contributions)
\begin{equation}
\label{kubo}
I_{\nu\sigma}^{BS}=i\int dt' \hspace{1mm}\theta\left(t-t'\right)\langle\left[\hat{H}^{
(0)}_{tun}(t'),\hat{I}_{\nu\sigma}^{BS,(0)}(t)\right]\rangle_{0},
\end{equation}
where the average $\langle ... \rangle_{0}$ is taken with respect to $\hat{H}^{HLL} + \hat{H}_g$.\\
In the above equation we have introduced the index $^{(0)}$ to indicate the time evolution of operators evaluated in absence of tunneling, i.e. with respect to $\hat{H}_{HLL}+ \hat{H}_g$ for the top edge and to $\hat{H}_{HLL}$ for the bottom one.
We remind the reader that in our model the external gate potential couples only to top edge electrons, see Fig.~\ref{giunzione1}.
In order to evaluate the expression in Eq.~(\ref{kubo}), it is sufficient to know the time evolution of fermionic operators, obtained from $\hat{H}_{HLL}+\hat{H}_g$. In particular we are interested only in their time-evolution close to the QPC positions at $x=\pm d$, since we are assuming local tunneling with a $\delta$- like spatial shape for the QPCs centered around $x=\pm d$.
The explicit derivation of the time-evolution of these operators is reported in Appendix \ref{ex_gate} for the case of a sinusoidal voltage $V(t)=V_{0} \cos(\omega t)$, and here we quote only the relevant results:
\begin{align}
\hat{\psi}_{R,\uparrow}(x,t)&=e^{iA_{1 \uparrow}\cos\left[\omega t-\theta_{1 \uparrow}(x)\right]}\hat{\psi}^{({ HLL})}_{R,\uparrow}(x,t),\label{new_fermionicR2}\\
\hat{\psi}_{L,\downarrow}(x,t)&=e^{iA_{1 \downarrow}\cos\left[\omega t-\theta_{1 \downarrow} (x)\right]}\hat{\psi}^{({HLL})}_{L,\downarrow}(x,t),\label{new_fermionicL2}
\end{align}
in the region close to the QPC1 (right) at $x=d$ and 
\begin{align}
\hat{\psi}_{R,\uparrow}(x,t)&=e^{iA_{2 \uparrow}(x) \cos\left[\omega t-\theta_{2 \uparrow}(x)\right]}\hat{\psi}^{({ HLL})}_{R,\uparrow}(x,t),\label{new_fermionicR}\\
\hat{\psi}_{L,\downarrow}(x,t)&=e^{iA_{2 \downarrow}(x) \cos\left[\omega t-\theta_{2 \downarrow}(x)\right]}\hat{\psi}^{({ HLL})}_{L,\downarrow}(x,t),\label{new_fermionicL}
\end{align}
in the region around the QPC2 (left) located at $x=-d$. In the above equations we have denoted with $\hat{\psi}_{r,\sigma}^{(HLL)}(x,t)$ the time-evolution of operators with respect to the bare $\hat{H}_{HLL}$, without the external gate.
The amplitudes and phases introduced just above are given by
\begin{align}
\label{ARL}
A_{2 \sigma}(x)&=\frac{eV_0}{2\omega}\sqrt{\alpha_{\sigma}^2(x)+\beta_{\sigma}^2(x)},\\
\theta_{2 \sigma}(x)&=\arctan{\frac{\alpha_{\sigma}(x)}{\beta_{\sigma}(x)}}+2m\pi \hspace{5mm}\left(\text{with } m \in \mathbb{N}\right),\\
A_{1 \uparrow/1 \downarrow}&=\frac{eV_0}{\omega}(1\pm K)\sin\left[\frac{\omega}{2u}\left(L_2+L_1\right)\right],\label{AR1}\\
\theta_{1 \sigma}(x)&=\frac{\omega}{2u}\left(L_2-L_1+2x\right),
\end{align} 
where the amplitudes $A_{1 \sigma}$ have no spatial dependence, contrary to what happens for $A_{2 \sigma}(x)$ which is affected by the gate voltage $V_{g}$.
The coefficients $\alpha_{\sigma}(x)$ and $\beta_\sigma (x)$ are connected to the geometry of the setup and are given by
\begin{subequations}
\label{rel00_giunzione1}
\begin{align}
\alpha_{\uparrow/\downarrow}(x)&=1-\cos{\left[\omega\left(\frac{L_2+L_1}{2u}\right)\right]}\cos{\left[\omega\left(\frac{x}{u}+\frac{L_2-L_1}{2u}\right)\right]}+\nonumber\\&\pm K \sin{\left[\omega\left(\frac{L_2+L_1}{2u}\right)\right]}\sin{\left[\omega\left(\frac{x}{u}+\frac{L_2-L_1}{2u}\right)\right]},\label{alpha00}\\
\beta_{\uparrow/\downarrow}(x)&=\sin{\left[\omega\left(\frac{L_2+L_1}{2u}\right)\right]}\cos{\left[\omega\left(\frac{x}{u}+\frac{L_2-L_1}{2u}\right)\right]}+\nonumber\\&\pm K \cos{\left[\omega\left(\frac{L_2+L_1}{2u}\right)\right]}\sin{\left[\omega\left(\frac{x}{u}+\frac{L_2-L_1}{2u}\right)\right]}.\label{beta00}
\end{align}
\end{subequations}
One can immediately note that in the non-interacting case, i.e. $K=1$, the amplitude $A_{1 \downarrow}=0$ vanishes. This will have significant implications on the pumped charge and heat currents as we will discuss below.\\
We recall that the tunneling is local at the QPC positions, with amplitude $h(x)=\Lambda [\delta(x-d) + \delta(x+d)]$, and we introduce the following series representation\cite{sharma01, sharma03}
\begin{align}
&h(x)e^{iA_{1 \sigma/ 2 \sigma}(x)\cos\left[\omega t-\theta_{1 \sigma/2 \sigma}(x)\right]}=\Lambda\sum\limits_{n=0}^{\infty}(i)^{n}\left(2-\delta_{n,0}\right)\cdot\nonumber\\&\cdot\delta(x\mp d)J_{n}(A_{1 \sigma/ 2 \sigma}(x))\cos\left[n\left(\omega t-\theta_{1 \sigma/ 2 \sigma}(x)\right)\right],\label{funzione_f}
\end{align}
where $J_n(x)$ is the Bessel function of $n$-th order. Inserting this relation into Eq.~(\ref{kubo}) one can obtain the expressions for the average pumped currents. We stress that the only contributions come from positions located at the QPCs $x=\pm d$. We thus introduce the simplified notations 
\begin{align}
A_{2 \sigma}(x=-& d)\equiv  A_{2 \sigma}
\qquad \theta_{1 \sigma}(x=d) \equiv \theta_{1 \sigma}~,\nonumber\\&
\qquad \theta_{2 \sigma}(x=-d) \equiv \theta_{2 \sigma}~,
\end{align}
and we obtain the dc components:
\begin{align}
I^{BS}_{N\sigma}&=2i\frac{\left|\Lambda\right|^2}{\left(\pi a\right)^2}\sum\limits_{n=1}^{\infty}J_n(A_{1 \sigma})J_n(A_{2 \sigma})\sin\left(n\left(\theta_{2 \sigma}-\theta_{ 1 \sigma}\right)\right)\cdot\nonumber\\\cdot&\sin\left(4k_{\rm F} d\right)\int d\tau \sin\left(n\omega \tau\right) P_{\gamma}\left(\tau-\frac{2d}{u}\right)P_{\gamma}\left(\tau+\frac{2d}{u}\right),
\label{particle_current}
\end{align}
for the average particle current per spin component, and
\begin{align}
&I^{BS}_{q\sigma}=i\frac{\left|\Lambda\right|^2}{\left(\pi a\right)^2}\frac{v_{\rm F}}{u}\sum\limits_{n=1}^{\infty}J_n(A_{1 \sigma})J_n(A_{2 \sigma})\sin\left(n\left(\theta_{2 \sigma}-\theta_{1 \sigma}\right)\right)\cdot\nonumber\\\cdot&\cos\left(4k_{\rm F} d\right)\int d\tau \sin\left(n\omega \tau\right)\Big( \partial_{\tau}P_{\gamma}\left(\tau-\frac{2d}{u}\right)P_{\gamma}\left(\tau+\frac{2d}{u}\right)+\nonumber\\&-P_{\gamma}\left(\tau-\frac{2d}{u}\right)\partial_{\tau}P_{\gamma}\left(\tau+\frac{2d}{u}\right)\Big),
\label{heat_current}
\end{align}
for the contribution associated to the heat current per spin component.
Notice that the $n=0$ contributions in the infinite sums cancel out and the series expressions present in Eqs.(\ref{particle_current})-(\ref{heat_current}) start from $n=1$. In the above equations we have introduced  $a$ the usual short-distance cut-off of HLL~\cite{martin05}, and $P_g(t)=e^{g\mathcal{W}(t)}$, where $\mathcal{W}(t)$ is the bosonic correlation function \cite{martin05, carrega11,carrega12, braggio12}
\begin{equation}
\label{boso_correlator}
\mathcal{W}(t)=\ln{\frac{\left|\Gamma\left(1+\frac{k_{{\rm B}} T}{\omega_{\rm c}}+i\ k_{{\rm B}}T t\right)\right|^2}{\Gamma^2\left(1+\frac{k_{{\rm B}}T}{\omega_{\rm c}}\right)\left(1+i\omega_{\rm c} t\right)}}.
\end{equation}
Here, $\omega_{\rm c}=v_{\rm F}/a$ the energy cut-off, which represents the highest energy scale of the problem. Note that this quantity can be related to the energy gap between bulk conduction and valence bands of the 2dTI.  The dimensionless parameter $\gamma$ depends on the interaction strength and reads
\begin{equation}
\gamma\equiv\frac{1}{2}\left(\frac{1}{K}+K\right).
\end{equation}
Performing the integrals present in Eqs.~(\ref{particle_current})-(\ref{heat_current}) (see Refs.~\onlinecite{chamon97} and Appendix~\ref{formule} for details) the average pumped currents can be written as
\begin{align}
{I^{BS}_{N\sigma}}&=4\left|\lambda\right|^2\sum\limits_{n=1}^{\infty}J_n(A_{1 \sigma})J_n(A_{2 \sigma})\sin\left(n\left(\theta_{2 \sigma}-\theta_{1 \sigma}\right)\right)\times\nonumber\\\times&\sin\left(4k_F d\right)\mathcal{H}\left(\gamma, \frac{2d n\omega}{u}, \frac{2d k_B T}{u}\right)\times\nonumber\\\times&\left[\tilde{\mathcal{P}}_{2\gamma}(n\omega)-\tilde{\mathcal{P}}_{2\gamma}(-n\omega)\right],\label{particle_current_H}\\
{I^{BS}_{q\sigma}}&=-v_{\rm F}\left|\lambda\right|^2\sum\limits_{n=1}^{\infty}J_n(A_{1 \sigma})J_n(A_{2 \sigma})\sin\left(n\left(\theta_{2 \sigma}-\theta_{1 \sigma}\right)\right)\times\nonumber\\\cdot&\cos\left(4k_F d\right)\Xi\left(\gamma,2d,\frac{n\omega}{u}, \frac{k_B T}{u}\right)\cdot\nonumber\\\times&\left[\tilde{\mathcal{P}}_{2\gamma}(n\omega)-\tilde{\mathcal{P}}_{2\gamma}(-n\omega)\right],\label{heat_current_Xi}
\end{align}
where $\lambda=\Lambda/(2\pi a)$ and $\mathcal{H}$ and $\Xi$ are modulating functions (see Appendix~\ref{formule} for the explicit expressions) and $\tilde{\mathcal{P}}_g(E)$ is the Fourier representation in energy domain of ${\mathcal{P}}_g(t)$. In passing we note that $\mathcal{H}$ and $\Xi$ do not depend on the voltage amplitude $V_0$, but this dependence is crucially present in the argument of the Bessel functions. 
Before discussing the behavior of the average pumped currents some comments are in order.
First of all, contrary to previous works\cite{ferraro13, chamon97}, here time-dependent tunneling amplitudes and phase shifts of the two QPCs are not introduced a priori in the system. Indeed, the solution of the microscopic model of the coupling to the external gate have lead to  the time-dependence of the tunneling amplitude and to the microscopic derivation of the phase shifts $\theta_{1 \sigma}$ and $\theta_{2 \sigma}$.
One can note that dc finite current signals strongly rely on quantum interference effects~\cite{giazottorev, ronetti16, vannucci15}, induced by the presence of the two constrictions in this geometry. Indeed, both currents in Eqs.~(\ref{particle_current_H})-(\ref{heat_current_Xi}) vanish in the limit $d\to0$.
From a physical point of view, each contributions in the sum over $n$ account for tunneling processes in which electrons absorb or emit $n$ photons. Therefore the tunneling amplitudes, for each QPCs, associated to these processes result weightened by the corresponding Bessel function $J_n(x)$ present in Eq.~(\ref{funzione_f}).\\
We also note that quantum interference effects are suppressed if the tunneling amplitude through the left ($x=-d$) QPC $A_{2 \sigma}$ or through the right ($x=d$) QPC $A_{1 \sigma}$ vanish $A_{1 \sigma/2 \sigma}\to0$, resulting in a vanishing pumped current signals. Finally we underline that the phases $\theta_{1 \sigma}$ and $\theta_{2 \sigma}$ acquired by electrons, have to be different in order to generate finite dc currents. This fact is consistent with the general prescription of the parametric pumping mechanism~\cite{giazotto11, thouless, brouwer,sharma01,sharma03}.
Once Eqs. \eqref{particle_current_H} and \eqref{heat_current_Xi} have been obtained, all stationary currents can be computed by using the relations in Eq.\eqref{charge_N} and Eq.\eqref{heatback1}.
\section{Results and discussion \label{results}}
In this section we will discuss our main results. We will mainly focus on the generation of finite heat current induced by the pumping mechanism in absence of external bias.
We investigate the net amount of heat current flowing through the 2dTI, studying the behavior of $I_q^{BS}$ as a function of various external parameters, inspecting also the role played by e-e interactions.\\
In the following all energies are rescaled with respect to the chemical potential $\mu$ (dimensionless quantities are thus indicated with a bar , i.e., $\overline{\omega}=\omega/\mu$, $\overline{V}_0 = e V_0/\mu$, $\overline{T}=k_{{\rm B}}T/\mu$, $\overline{\omega}_{{\rm c}}=\omega_{{\rm c}}/\mu$).
For sake of convenience, we also introduce
\begin{equation}
\eta= d k_{{\rm F}}= d\frac{\mu}{v_{\rm F}}~,
\end{equation}
a dimensionless parameter directly connected to the presence of quantum interference effects, as it appears in the expressions for the pumped current Eqs.~(\ref{particle_current_H})-(\ref{heat_current_Xi}) and is linearly proportional to the distance between the two QPCs.\\
We restrict the analysis to a physically reasonable range of parameters, remaining as close as possible to experimentally accessible values. We fix the chemical potential to $\mu=3$~meV and the temperature to $T=300$~mK ($\overline{T}=0.01$).
The energy cut-off, which represents the largest energy scale involved, is related to the typical energy bulk gap in 2dTI. Recent proposal have predicted very wide bulk gap for 2dTI in topological materials with strong spin-orbit coupling~\cite{reis16, zhang16, zhou15}. Recent measurement on novel devices have reported bulk gap values up to $\sim 800$~meV~\cite{reis16}. Here we fix the energy cut-off to $\overline{\omega}_{{\rm c}}=200$.\\
Concerning the characteristic parameters of the external gate, here we fix $\overline{\omega}=0.1$ which corresponds to a drive frequency of $\nu=\omega/(2\pi)\sim 75$~GHz. We vary the gate amplitude in the range $0\leq \overline{V}_0 \leq 20$. In this case, we can investigate a large range of parameters, including the non-linear regime $eV_0/\omega \gg 1$ where interesting features appear.\\
\begin{figure}[htb]
\centering
\includegraphics[width=.45\textwidth]{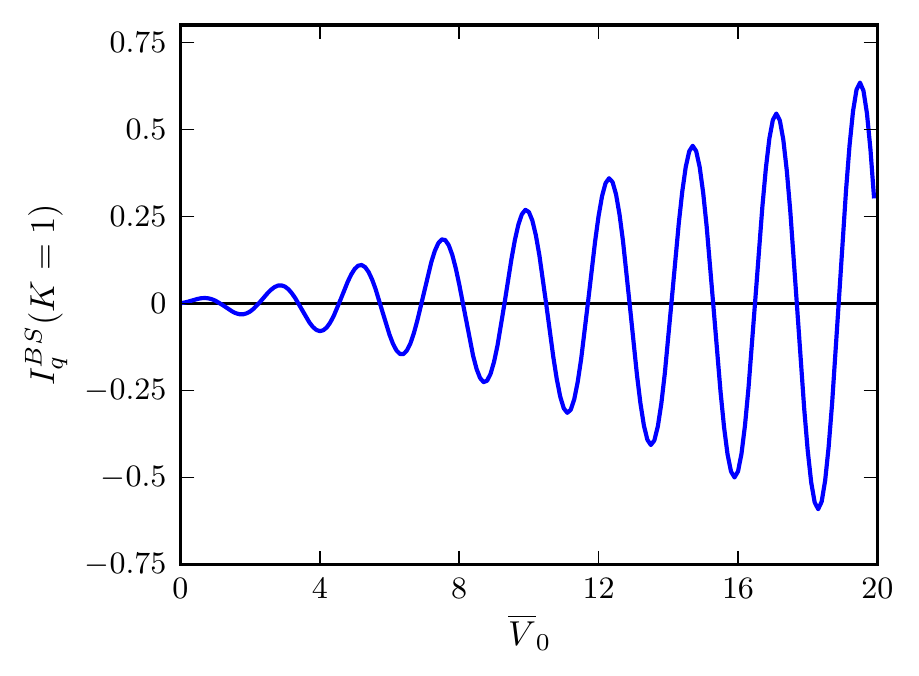}\quad
\caption{(Color online) Average pumped heat current $I_q^{BS}$ (in units of $10^{-3}|\lambda^2|/\overline{\omega}_{{\rm c}}$) as a function of the external gate voltage $\overline{V}_0= e V_0/\mu$ in the non-interacting case ($K=1$). Drive frequency and temperature are fixed to $\overline{\omega}=0.1$ and $\overline{T}=0.01$. Other parameters are $\eta=1.5$, $L_1=0.2 d$, $L_2=2 d$ and the energy cut-off is set to $\overline{\omega}_{{\rm c}}=200$.
As discussed in the text, finite heat current is generated by pumping and for $K=1$ it is only carried by $\uparrow$ electrons, resulting in a flow of polarized heat current.
}
\label{fig2}
\end{figure}
Let us start considering the non-interacting case ($K=1$). Interestingly enough we get $A_{1 \downarrow}=0$ which, in turn, lead to the vanishing of the heat current component associated to $\downarrow$ electrons $I_{q\downarrow}^{BS}=0$. This fact lead to the conclusion that, in the non-interacting case, a finite heat current is produced and is associated {\it only} to a single spin species $\sigma=\uparrow$
\begin{equation}
\lim_{K=1}
I_{q}^{BS}= I_{q \uparrow}^{BS}.
\end{equation}
Therefore this setup can allow for the generation of fully polarized heat current.
Note that the vanishing of the component associated to $\downarrow$ electrons with $A_{1 \downarrow}\to0$ is intimately connected to the spin-momentum locking property of 2dTI.
In the end this is also connected to the chosen geometry. Indeed if the external gate is placed on top of the other QPC, this would imply $A_{2 \uparrow}\to 0$ with an opposite polarization of the pumped heat current.
Indeed, physically, the vanishing of spin $\downarrow$ component can be understood in terms of tunneling paths of electrons. When $L,\downarrow$ electrons reach the right QPC, they have not increased or reduced their energy yet by passing through the gate. Therefore, each contribution to transport that could arise from $L,\downarrow$ tunneling at $x=d$ is exactly compensated by contribution $R,\downarrow$ tunneling at the right QPC.
Only the tunneling path through the other QPC is due to $\downarrow$ electrons and interference effects would completely vanish.\\
In Fig.~\ref{fig2} we show the average pumped heat current $I_{q}^{BS}=I_{q\uparrow}^{BS}$ in the non-interacting case, where an oscillating behavior is clearly visible as a function of the gate voltage $\overline{V}_0$.
The latter is due to the Bessel functions $J_{n}\left(A_{1 \uparrow}\right)$ and $J_{n}\left(A_{2 \uparrow}\right)$ present in Eq.~(\ref{heat_current_Xi}).
In order to understand this oscillating behavior, it is instructive to look at the limit of small $\eta\overline{\omega}$ of $A_{1 \uparrow}$ (as it is in our case with the chosen parameters) 
\begin{equation}
\lim_{\eta\overline{\omega} < 1}A_{1 \uparrow}\to \overline{V}_0 \eta \frac{L_1 + L_2}{d}~,
\end{equation}
proportional to $\overline{V}_0$.
One can see that for $\eta (L_1+L_2)/d>1$, as it is in our case, maxima and minima of $J_{n}\left(A_{1 \uparrow}\right)$, as a functions of the gate amplitude, are several in the considered range of $\overline{V}_0$. This basically would explain the oscillating pattern present in Fig.~\ref{fig2} for the polarized heat current in the non-interacting case. Indeed the other $J_n(A_{2 \uparrow})$ present a slower oscillating behavior, since $A_{2\uparrow}$ is linked to $L_2 - d$ always smaller than $L_1+L_2$. We underline that the same qualitative features are present also for higher values of $\overline{\omega}\eta$, where in general the full expression of $A_{1 \uparrow}$ governs the observed pattern. Finally, the magnitude of $I_{q}^{BS}$  is related to the weigth function present in Eq.~(\ref{heat_current_Xi}) and to the product of Bessel functions itself. The latter are responsible of the increasing magnitude for increasing gate voltage $\overline{V}_0$ in the chosen range of parameters, where Bessel functions of higher order $n$ become relevant.\\
The presence of e-e interactions strongly modifies the above picture, as we now discuss.
\begin{figure}[htb]
\centering
\includegraphics[width=.36\textwidth]{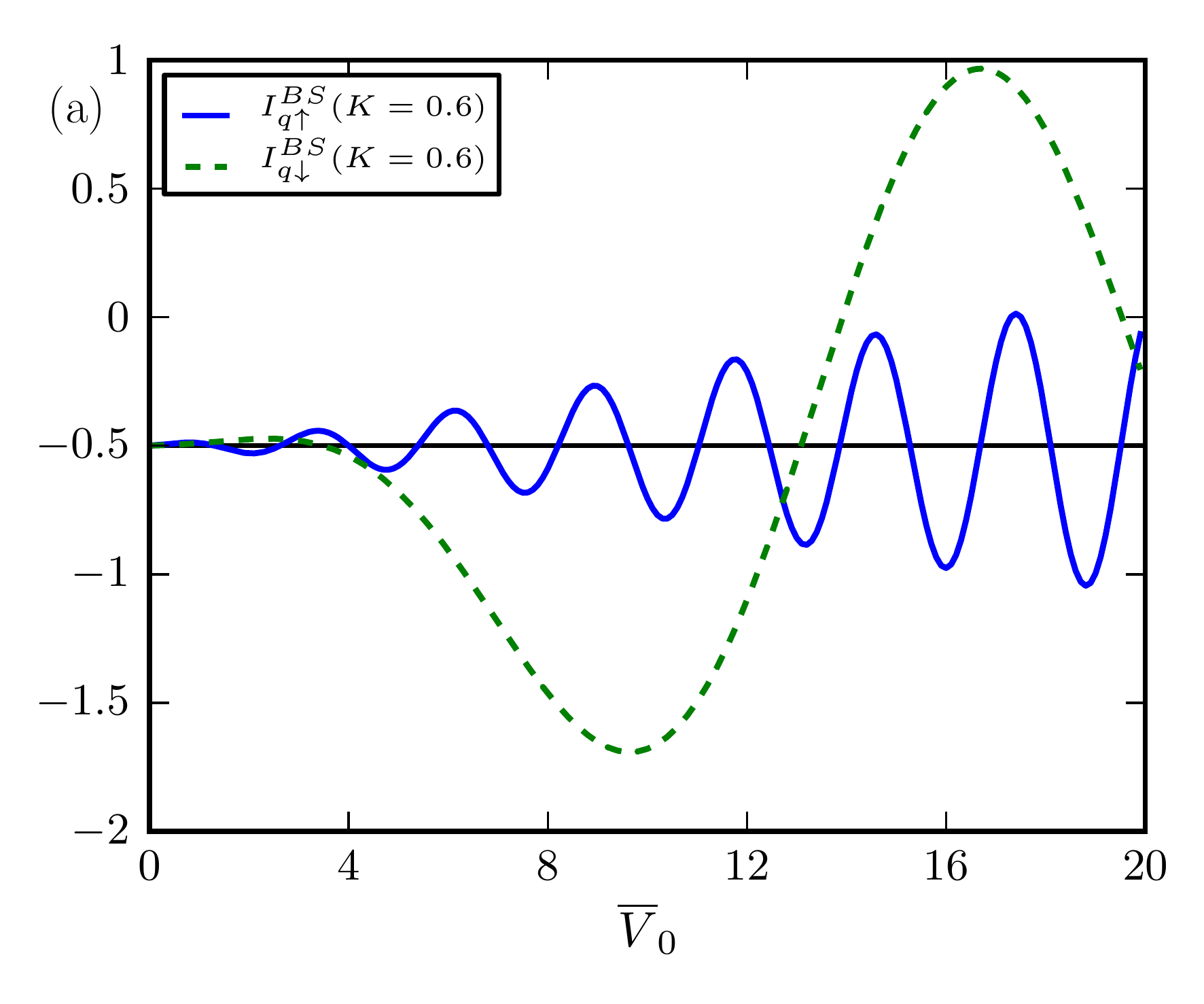}
\includegraphics[width=.36\textwidth]{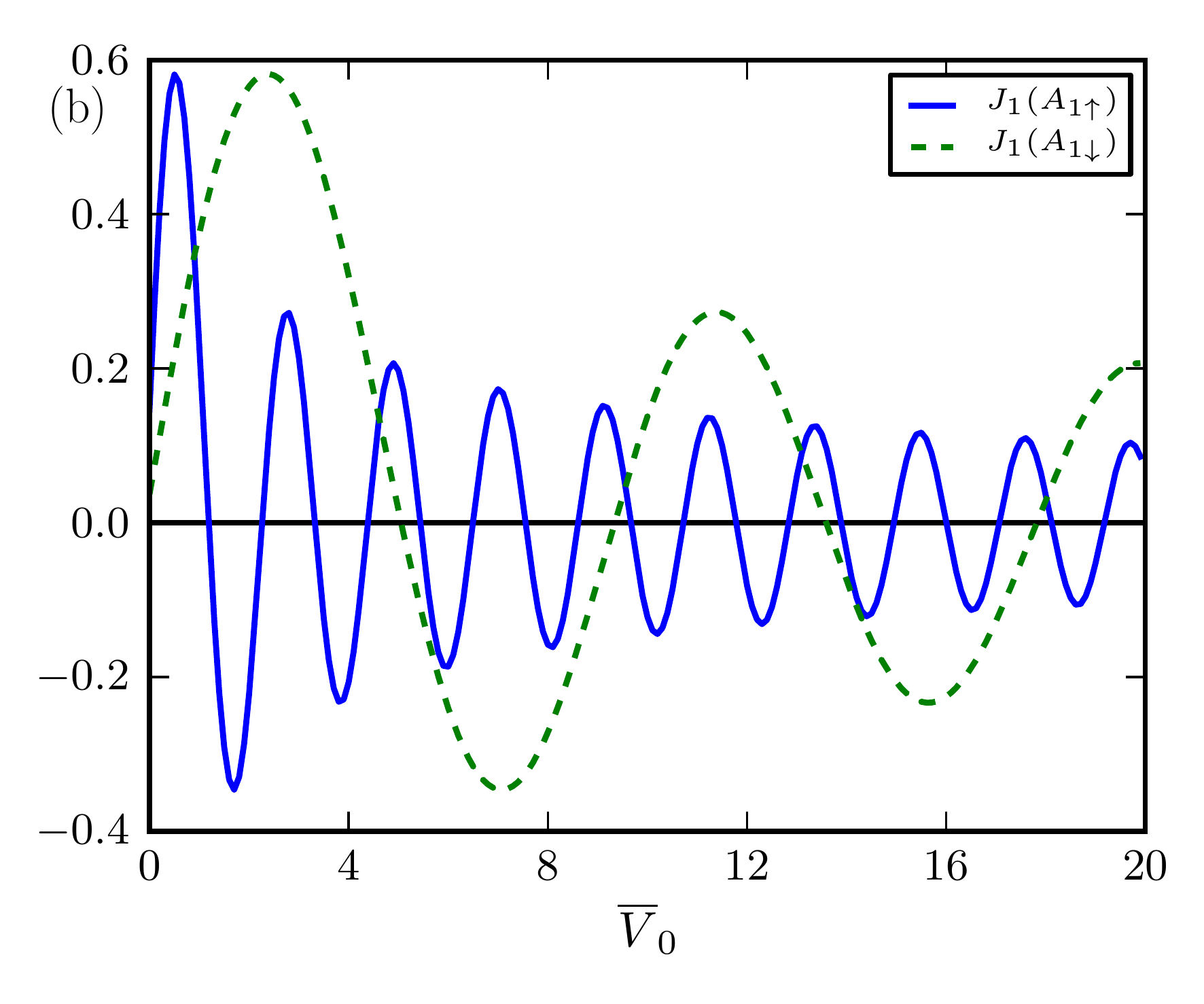}
\includegraphics[width=.36\textwidth]{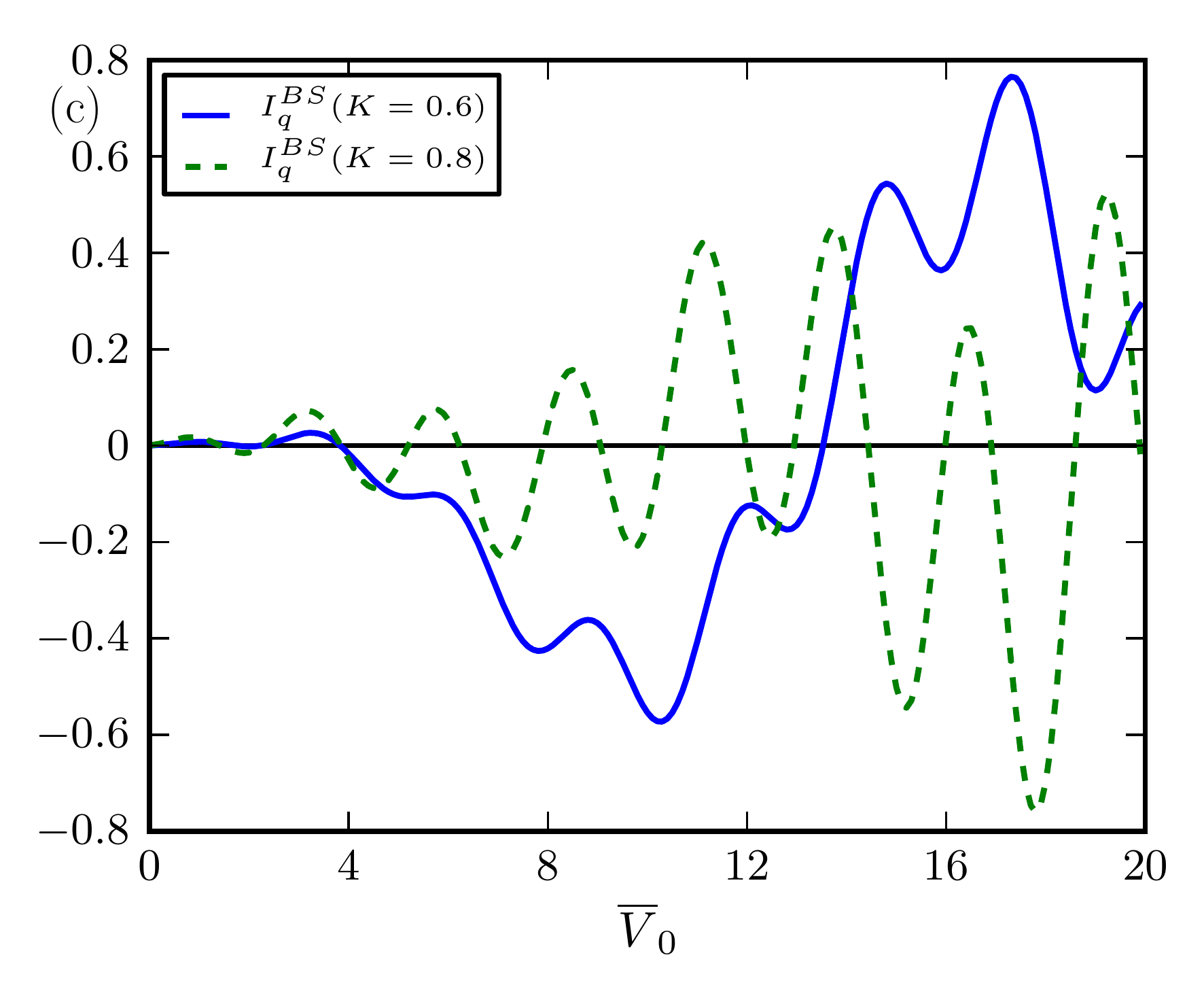}
\caption{(Color online) Average pumped heat current in presence of e-e interactions ($K<1$).
Panel a) shows the spin up $I_{q \uparrow}^{BS}$ (blue solid line) and spin down $I_{q \downarrow}^{BS}$ (green dashed line) components of the average heat current for $K=0.6$.
Panel b) represents the behavior of the more relevant Bessel functions with $n=1$ with interaction strength $K=0.6$.
Panel c) shows the heat current $I_q^{BS}$ (in units of $10^{-3}|\lambda^2|/\overline{\omega}_{{\rm c}}$) as a function of the gate voltage $\overline{V}_0=e V_0/\mu$. Blue solid line and green dashed curve correspond to $K=0.6$ and $K=0.8$ respectively.
Other parameters are the same as in Fig.~\ref{fig2}.
}
\label{fig3}
\end{figure}
The interacting case is reported in Fig.~\ref{fig3}. The first important difference with respect to the non-interacting case is that now both spin components contribute to the average pumped heat current $I_q^{BS}=I_{q \uparrow}^{BS}+I_{q \downarrow}^{BS}$, and thus the heat current is not polarized anymore, see in particular Fig.~\ref{fig3}b).\\ The behaviours of $I_{q \uparrow}^{BS}$ and $I_{q \downarrow}^{BS}$ can be ascribed qualitatively to the difference between the Bessel functions  $J_{n}\left(A_{1 \uparrow}\right)$ and  $J_{n}\left(A_{1 \downarrow}\right)$. These functions are plotted in Fig.~\ref{fig3}b) versus the gate voltage $\overline{V}_0$.
As one can see $J_{n}\left(A_{1 \uparrow}\right)$ displays more oscillations than $J_{n}\left(A_{1 \downarrow}\right)$.
Indeed, looking at Eq. \eqref{AR1} one can note that $A_{1 \downarrow}=\epsilon A_{1 \uparrow}$, with $\epsilon=\frac{1-K}{1+K}$, which is $\epsilon<1$ for all interaction strengths in the interval $0<K\le 1$.
Therefore, the number of zeros of $J_{n}\left(A_{1 \downarrow}\right)$ are reduced of a factor $\epsilon$ with respect to $J_{n}\left(A_{1 \uparrow}\right)$as a function of $\overline{V}_0$. We underline that this factor $\epsilon$, related to interaction parameter $K$, arises as a consequence of the particular form for the phases of the bosonized expression for $\hat{\psi}_{R,\uparrow}$ and $\hat{\psi}_{L,\downarrow}$ in Eqs. \eqref{boso_sol} and is a peculiar property of HLL.\\
The difference between $A_{1 \uparrow}$ and $A_{1\downarrow}$ are also reflected in the different amplitudes of spin components of the heat current. Summing over higher $n$-th order of the Bessel functions and considering the faster oscillation  related to $A_{1 \uparrow }$, the amplitude of $I_{q \uparrow}^{BS}$ decreases with respect to $I_{q\downarrow}^{BS}$, see Fig.~\ref{fig3}a).
This trend is reflected in the total heat current, i.e. the sum of the two spin contributions, which is thus dominated by the slower oscillation associated to $I_{q \downarrow}^{BS}$, as one can see in Fig.~\ref{fig3}c). This fact is clearer if one consider stronger interactions (see the green dashed line for $K=0.6$), where the period of oscillations is dominated by the slow spin $\downarrow $ component, presenting some beats due to the modulation with the fast oscillation associated to $I_{q\uparrow}^{BS}$. Due to all these different features, the pumped heat current can be also used as a sensitive probe of the presence of e-e interactions. Indeed, as we have shown the heat current is no more fully polarized and presents a different pattern of oscillations, with characteristic modulations related to the interaction strength.\\
\begin{figure}[h]
\centering
\includegraphics[width=.45\textwidth]{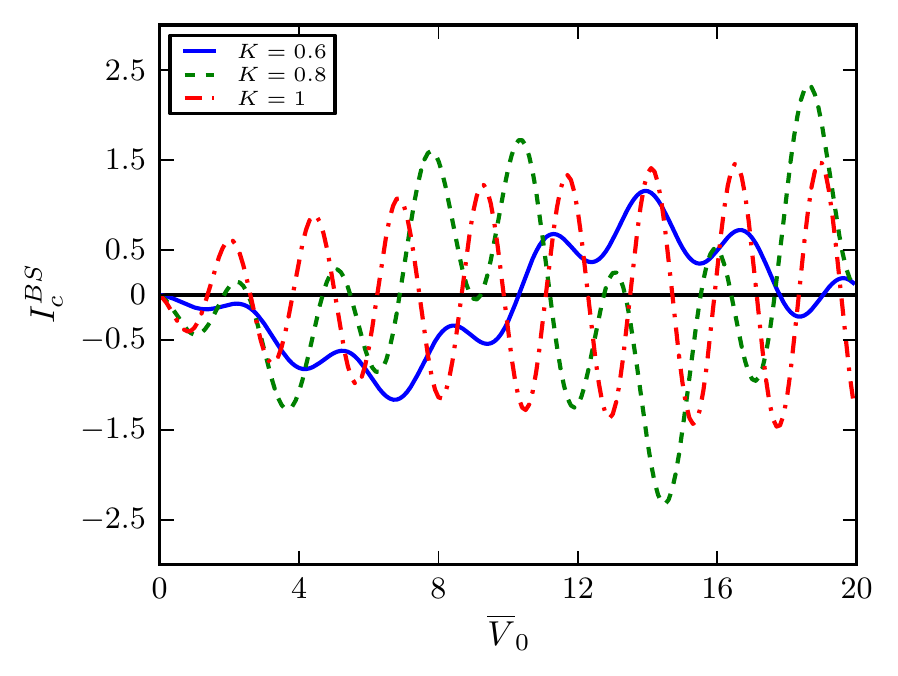}
\caption{(Color online) Pumped charge current $ I^{BS}_{c} $, in units of
$4\cdot 10^{-4} e\frac{\left|\lambda\right|^2}{\omega}$, as a function of the gate amplitude $\overline{V}_0$. Different curves correspond to different interaction strengths $K=0.6$ (blue solid line), $K=0.8$(green dashed line), $K=1$ (red line). Other parameters are the same as in Fig.~\ref{fig2}.
}
\label{fig4}
\end{figure}
A similar behavior is expected also for the average charge current. Indeed the same pumping mechanism would lead to a finite charge current signal. This is reported in Fig.~\ref{fig4}, where $I_c^{BS}$ is plotted as a function of the gate voltage $\overline{V}_0$  for different interaction strength.
It turns out that in the non-interacting case also pumped charge current is fully polarized and carried solely by spin $\uparrow$ electrons. The presence of interactions induces the same qualitative changes discussed for the heat current and  the associated oscillations can be explained in full analogy to what happens for $I_q^{BS}$.
\begin{figure}[h]
\centering
\includegraphics[width=.40\textwidth]{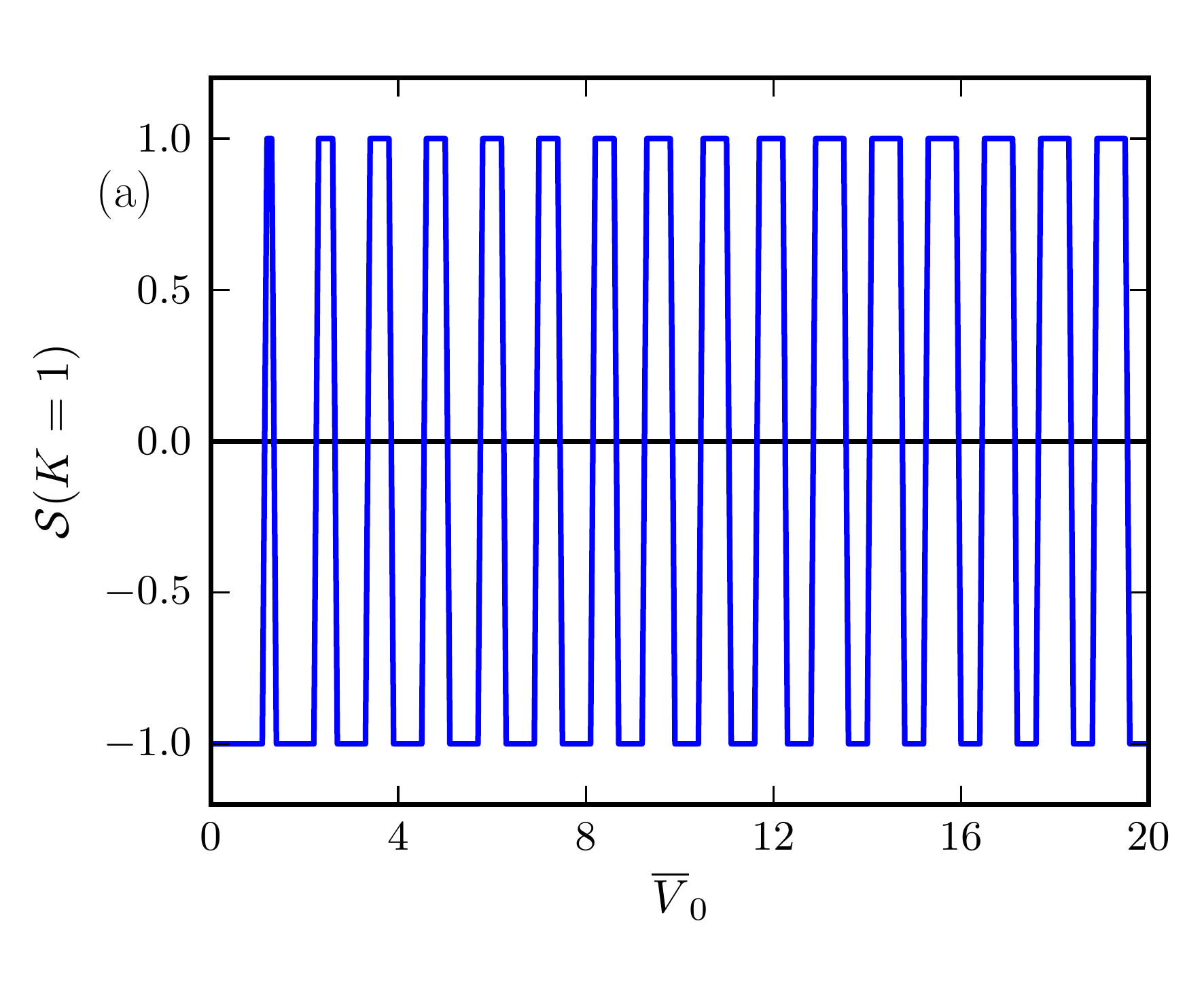}\quad
{
\includegraphics[width=.40\textwidth]{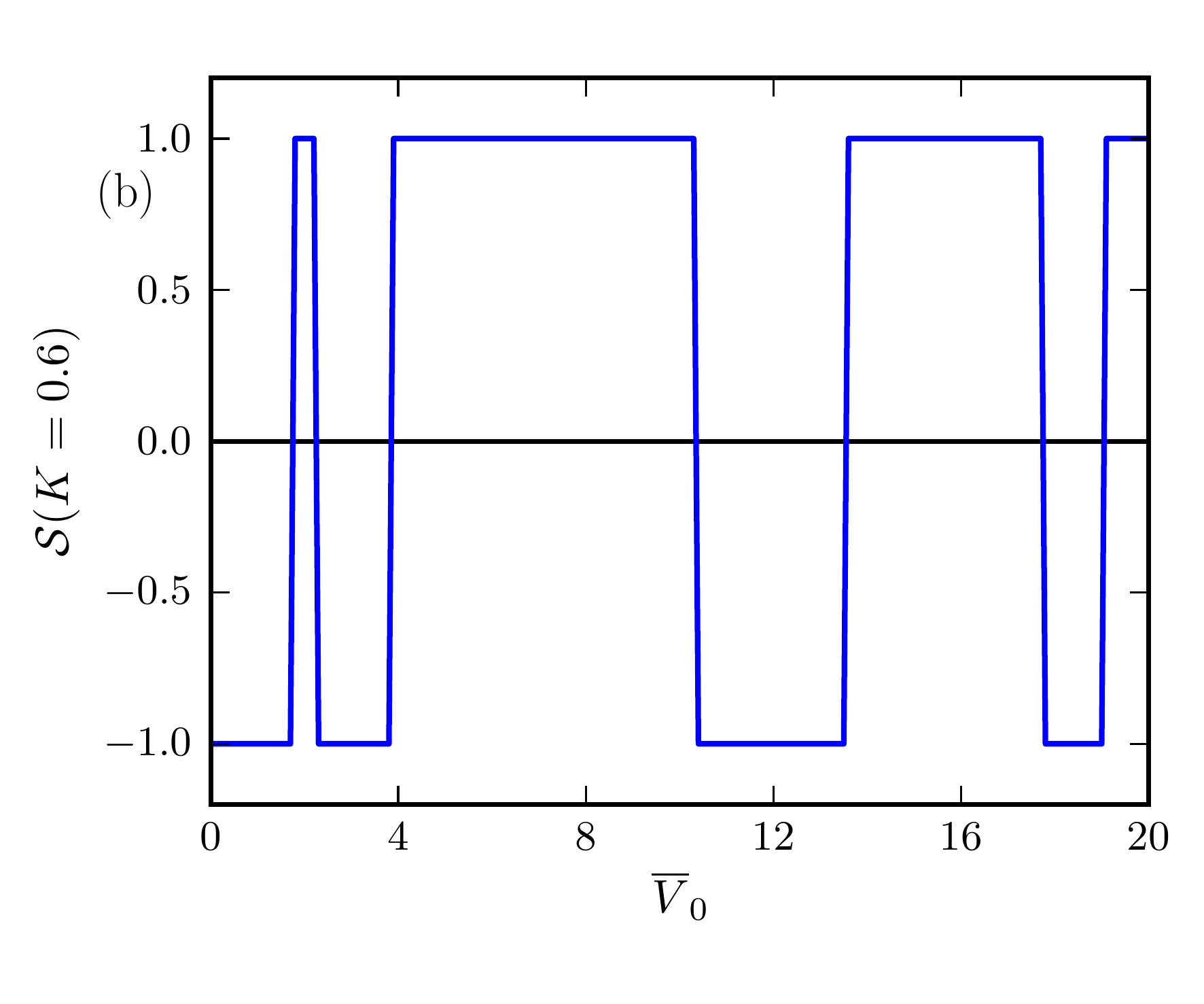}} 
\caption{(Color online) Sign of the product of charge current $ I^{BS}_{c} $ and heat current $ I^{BS}_{ q} $, as a function of the gate amplitude $\overline{V}_0$. In the upper panel $K=1$(non-interacting case) and in the lower panel $K=0.6$. Other parameters are the same as in Fig.~\ref{fig2}.}
\label{fig5}
\end{figure}
It is interesting, however, to look at the sign of the product of the two pumped currents, i.e.
\begin{equation}
\label{sign}
{\cal S}(\overline{V}_0, K)\equiv \frac{I_c^{BS} I_q^{BS}}{|I_c^{BS}||I_q^{BS}|}~,
\end{equation}
as a function of the gate amplitude $\overline{V}_0$ as shown in Fig.~\ref{fig5}.
Here we show the non-interacting case ($K=1$) in the upper panel and $K=0.6$ in the lower panel.
In general, the fact that this quantity has not a definite sign indicates that by varying the gate voltage one can achieve regimes in which charge and heat currents flows in opposite directions.
The difference in the sign of charge and heat current, as a function of the gate voltage, can be related to the modulating functions $\mathcal{H}$ and $\Xi$, which weight the contribution of each photo-assisted tunneling occuring at energy $n\omega$. In the presence of interactions, when $A_{1\downarrow}\ne 0$, the lower number of oscillations of $J_{n}\left(A_{1 \downarrow}\right)$ with respect to $J_{n}\left(A_{1 \uparrow}\right)$ allows to better distinguish the effect of processes of each order $n$ on the sign of charge and heat currents. On the other hand, for $K=1$, only $J_{n}\left(A_{1 \uparrow}\right)$ remains and brings into each $n$-th process many oscillations, thus resulting in a strong reduction of the modulating function.
In the interacting case (see the lower panel in Fig.~\ref{fig5}) the regions with a definite sign of ${\cal S}$ as a function of $\overline{V}_0$ become wider with respect to the non-interacting one. Here charge and heat flow in the same - or opposite - directions for larger range of the external gate voltage $\overline{V}_0$.
\section{Conclusions \label{conclusions}}
In this work we have investigated a double quantum point contact geometry in a 2dTI. Based on this, we have proposed a setup in which a pumping mechanism can generate finite heat and charge currents without external dc bias or thermal gradient. The pumping mechanism relies on the presence of an external gate potential placed on top of one of the two constrictions which acts on the system with an external ac field.
We have developed a microscopic model of the coupling between the external gate and the electrons of the edge states of the 2dTI.
We have therefore evaluated the average heat and charge currents pumped into the system as a function of various parameters, taking also into account the presence of e-e interactions.
In the non-interacting case, this setup generates finite heat current which is carried solely by a single spin species, resulting into a fully polarized heat flow.
The presence of e-e interactions strongly modifies the behavior of pumped currents, which are no more fully polarized since both spin components now contribute to the net heat flow.
Moreover, looking at the behavior of the heat current as a function of the gate voltage, one can distinguish between different pattern of oscillations. The latter are connected to the interaction strength, thus allowing to use this setup as a tool to identify the presence of e-e interactions. Finally we have studied the sign of the product of heat and charge pumped currents, showing that one can reach situation in which the two quantities flow in the same or in the opposite directions, by varying the external gate voltage.

\begin{acknowledgments}
F.R., M.C., and M.S. acknowledge the support of the MIUR-FIRB2013 -- 
Project Coca (Grant No.~RBFR1379UX) and the COST Action MP1209. M.C. wish to thank A. Braggio, F. Taddei, and L. Arrachea for fruitful 
discussion. People from the Marseille group would like to acknowledge the support of Grant No. ANR-2010-BLANC-0412 (``1 shot'') and of 
ANR-2014-BLANC ``one shot reloaded''. This work was 
carried out in the framework of Labex ARCHIMEDE Grant No. 
ANR-11-LABX-0033 and of A*MIDEX project Grant No. ANR-11-IDEX-0001-02, 
funded by the ``investissements d'avenir'' French Government program 
managed by the French National Research Agency (ANR).

\end{acknowledgments}
\appendix
\section{Zero order currents \label{zero}}
This Appendix is devoted to the evaluation of the average currents in the absence of tunneling, taking into account the presence of the time-dependent external gate $V_g(x,t)$. In particular we are interested in average quantities in the dc limit or stationary regime. Here, we demonstrate that, in absence of tunneling events (i.e. without QPCs), all average currents vanish.\\
Since the gate potential couples only to electrons belonging to the top edge, we do not consider here operators related to the bottom edge.
Charge and energy densities associated to the top edge (denoted with the index $^{(1)}$) can be written as
\begin{subequations}
\label{density_ops}
\begin{align}
\hat{\rho}_{c}^{(1)}(x,t)&=-e\left(\hat{\rho}_{R,\uparrow}(x,t)+\hat{\rho}_{L,\downarrow}(x,t)\right),\\
\hat{\rho}_{q}^{(1)}(x,t)&=v_{\rm F}\left[\left(\hat{\rho}_{R,\uparrow}(x,t)\right)^2+\left(\hat{\rho}_{L,\downarrow}(x,t)\right)^2\right].
\end{align}
\end{subequations}
The corresponding currents can be defined by using the generalized continuity equation ($\nu=c,q$)
\begin{equation}
\label{zeroorderRL}
\partial_t\hat{\rho}^{(1)}_{\nu}(x,t)+\partial_x\hat{I}^{(1)}_{ \nu}(x,t)=0.
\end{equation}
Density operators, expressed in terms of bosonic fields, are given by
\begin{widetext}
 \begin{align}
\hat{\rho}_{R,\uparrow}(x,t)&=\frac{1}{2}\Bigg[\left(\frac{1}{\sqrt{K}}+\sqrt{K}\right)\partial_{x}\hat{\phi}_{+}^{(1)}(x,t)\left(\frac{1}{\sqrt{K}}-\sqrt{K}\right)\partial_{x}\hat{\phi}_{-}^{(1)}(x,t)\Bigg],\\
\hat{\rho}_{L,\downarrow}(x,t)&=-\frac{1}{2}\Bigg[\left(\frac{1}{\sqrt{K}}+\sqrt{K}\right)\partial_{x}\hat{\phi}_{-}^{(1)}(x,t)+\left(\frac{1}{\sqrt{K}}-\sqrt{K}\right)\partial_{x}\hat{\phi}_{+}^{(1)}(x,t)\Bigg].
\end{align}
\end{widetext}
We now use the results found in Appendix \ref{ex_gate}, where the time-evolution of operators $\hat{\phi}_{\pm}^{(1)}(x,t)$ in presence of the external gate has been calculated.
Given Eqs. \eqref{bosogate1} and \eqref{bosogate2}, we can calculate average currents in absence of tunneling events, in the dc limit.
After performing a thermal average with respect to $\hat{H}_{ HLL}+\hat{H}_g$, charge and heat currents are evaluated in the long-time limit (dc regime), using \eqref{zeroorderRL}
\begin{widetext}
\begin{align}
{I}_{c}^{(1)}(x,t) &=-e K\sum_{\zeta=+,-}\Bigg\{\theta\left(\zeta\left(x+L_2\right)\right)V\left(t-\zeta\frac{x+L_2}{u}\right)-\theta\left(\zeta\left(x-L_1\right)\right)V\left(t-\zeta\frac{x-L_1}{u}\right)\Bigg\},\label{Charge_zero}
\end{align}
for charge contribution,
\begin{align}
I_{q}^{(1)}(x,t)=\frac{e^2}{2u}&\left(K+K^2\right)\sum_{\zeta=+,-}\zeta\Bigg\{\theta\left(\zeta\left(x+L_2\right)\right) V^2\left(t-\zeta\frac{x+L_2}{u}\right)+\theta\left(\zeta\left(x-L_1\right)\right)V^2\left(t-\zeta\frac{x-L_1}{u}\right)+\nonumber\\&-2\theta\left(\zeta\left(x+L_2\right)\right)\theta\left(\zeta\left(x-L_1\right)\right) V\left(t-\zeta\frac{x+L_2}{u}\right)V\left(t-\zeta\frac{x-L_1}{u}\right)\Bigg\},\label{Heat_zero}
\end{align}
\end{widetext}
for heat current.\\
The expression in Eq. \eqref{Charge_zero} is a periodic function of time and it is linear in the time-dependent gate potential. Since the average over one period of $V(t)$ is zero,  it is easy to realize that charge current has no dc contribution in the absence of tunneling.\\
On the other hand, heat current is quadratic in the gate potential, with a non-trivial spatial dependence. In order to calculate its contribution, we thus consider two different regions with $x < -L_2$ and $x>L_1$.
In the first region ($x<-L_2$), the heat current, averaged over one period, reads 
\begin{equation}
I_{q\mathcal{L}}=-\frac{e^2V_0^2 }{2}\left(K+K^3\right)\left(1-\cos\left(\omega\frac{L_2+L_1}{u}\right)\right).
\end{equation}
while in the other region ($x>L_1$) one has 
\begin{equation}
I_{q\mathcal{R}}=\frac{e^2V_0^2 }{2}\left(K+K^3\right)\left(1-\cos\left(\omega\frac{L_2+L_1}{u}\right)\right).
\end{equation}
It is worth to note that $I_{q\mathcal{R}}$ corresponds to the dc heat current flowing into the right reservoir, while $-I_{q\mathcal{L}}$ represents the dc heat current flowing into the left reservoir. 
Since these two contributions are equal,  the net heat current globally flowing along the 2dTI is zero.
This means that the heat current introduced by the pumping mechanism, without tunneling events and backscattering, is partitioned on two equal parts one to the left and one to the right, resulting  into a zero net contribution to the pumped heat flow.

\section{Presence of an external gate \label{ex_gate}}
This Appendix is devoted to the derivation of the time-evolution of operators in presence of a time-dependent external gate $V_g(x,t)$.
As stated in the main text, we focus on a periodic time-dependent gate potential, whose explicit time-dependence is
\begin{equation}
\label{eqyt}
V(t)=V_0 \cos\left(\omega t\right).
\end{equation}
We start by evaluating the time-evolution of the operators $\hat{\phi}_{r,\sigma}(x,t)$ in presence of the external gate.
The Hamiltonian associated to the gate potential in Eq.~\eqref{Ham_gate_giunzione1} can be rewritten in terms of chiral bosonic fields as
 \begin{equation}
\label{Ham_gate_chirale}
\hat{H}_{{ g}}=e\sqrt{K}\int\limits V_g(x,t)\left[:\partial_x\hat{\phi}_{+}^{(1)}(x):-:\partial_x\hat{\phi}_{-}^{(1)}(x):\right].
\end{equation}
The equation of motion for the bosonic fields associated to the top edge $\hat{\phi}_{\zeta,}^{(1)}(x,t)$ (in the presence of $\hat H_ {HLL}+\hat H_g$) is
\begin{equation}
\label{eq_moto_gate2}
\partial_{t}\hat{\phi}_{\zeta}^{(1)}(x,t)+\zeta u \partial_x \hat{\phi}_{\zeta}^{(1)}(x,t)=-e\sqrt{K}\hspace{0.5mm}V_g(x,t),
\end{equation}
with the corresponding solution: 
\begin{widetext}
\begin{align}
\label{final_sol}
&\hat{\phi}_{\zeta}^{(1)}(x,t)=\hat{\phi}^{(1,{ HLL})}_{\zeta}(x-\zeta u t)-e\hspace{1mm}\sqrt{K}\int\limits_{0}^{t}V_g\left(x-\zeta u(t-t'),t'\right)dt'.
\end{align} 
\end{widetext}
Here, we have denoted with $\hat{\phi}^{(1,{ HLL})}_{\zeta}(x-\zeta ut)$ the solution in the absence of the external gate. 
Recalling the explicit form of the gate potential in Eq.~\eqref{ham_gate}, we have 
\begin{widetext}
\begin{align}
\Delta\hat{\phi}^{(1)}_{+}(x,t)&=-e\sqrt{K}\Bigg[\theta\left(-L_2<x<L_1\right)\int\limits_{t-\frac{L_2+x}{u}}^{t}dt'V\left(t'\right)+\theta\left(x>L_1\right)\int\limits_{t-\frac{-L_1+x}{u}}^{t-\frac{L_2+x}{u}}dt'V\left(t'\right)\Bigg]\label{bosogate1}\\\Delta\hat{\phi}^{(1)}_{-}(x,t)&=-e\sqrt{K}\Bigg[\theta\left(-L_2<x<L_1\right)\int\limits_{t+\frac{-L_1+x}{u}}^{t}V\left(t'\right)dt'+\theta\left(x<-L_2\right)\int\limits_{t+\frac{-L_1+x}{u}}^{t+\frac{L_2+x}{u}}V\left(t'\right)dt'\Bigg],\label{bosogate2}
\end{align}
\end{widetext}
where $\Delta\hat{\phi}^{(1)}_{\pm}(x,t)=\hat{\phi}^{(1)}_{\pm}(x,t)-\hat{\phi}^{(1,{\rm HLL})}_{\pm}(x,t)$. Recalling the explicit time-dependence in Eq.~(\ref{eqyt}), the expression in Eqs.~(\ref{bosogate1})-(\ref{bosogate2})
become
\begin{widetext}
\begin{align}
\Delta\hat{\phi}^{(1)}_{+}(x,t)=-\frac{eV_0}{\omega}\sqrt{K}\Bigg\{\theta\left(-L_2<x<L_1\right)\left[\sin\left(\omega t\right)-\sin\left[\omega\left(
t-\frac{x+L_2}{u}\right)\right]\right]+\nonumber\\+\theta\left(x>L_1\right)\left[\sin\left[\omega\left( t-\frac{x+L_2}{u}\right)\right]-\sin\left[\omega\left(
t-\frac{x-L_1}{u}\right)\right]\right]\Bigg\},\label{bosogate1a}\\\Delta\hat{\phi}^{(1)}_{-}(x,t)=-\frac{eV_0}{\omega}\sqrt{K}\Bigg\{\theta\left(-L_2<x<L_1\right)\left[\sin\left(\omega
t\right)-\sin\left[\omega\left( t+\frac{x-L_1}{u}\right)\right]\right]+\nonumber\\+\theta\left(x<-L_2\right)\left[\sin\left[\omega\left( t+\frac{x+L_2}{u}\right)\right]-\sin\left[\omega\left(
t+\frac{x-L_1}{u}\right)\right]\right]\Bigg\}.\label{bosogate2a}
\end{align}
\end{widetext}
\subsection{Time-evolution of fermionic operators}
\label{app:fermion}
We now evaluate the time-evolution of fermionic operators $\hat{\psi}_{R,\uparrow}(x,t)$ and $\hat{\psi}_{L,\downarrow}(x,t)$.
We recall that these operators are related to bosonic fields by the bosonization identity \cite{giamarchi03, miranda03}
\begin{align}
\hat{\psi}_{R,\uparrow/L,\downarrow}(x,t)&\sim e^{\frac{i}{2}\Bigg[\left(\frac{1}{\sqrt{K}}\pm\sqrt{K}\right)\hat{\phi}^{(1)}_{+}(x,t)+\left(\frac{1}{\sqrt{K}}\mp\sqrt{K}\right)\hat{\phi}^{(1)}_{-}(x,t)\Bigg]}. \label{boso_sol}
\end{align}
By writing $\hat{\phi}^{(1)}_{\zeta}(x,t)=\hat{\phi}^{(1,{HLL})}_{\zeta}(x,t)+\Delta \hat{\phi}^{(1)}_{\zeta}(x,t)$, Eq. \eqref{boso_sol} can be recast
as
\begin{widetext}
\begin{align}
\hat{\psi}_{R,\uparrow/L,\downarrow}(x,t)=\exp\Bigg\{-i\Bigg[\left(\frac{1\pm K}{2\sqrt{K}}\right)\Delta\hat{\phi}^{(1)}_{+}(x,t)\left(\frac{1\mp
K}{2\sqrt{K}}\right)\Delta\hat{\phi}^{(1)}_{-}(x,t)\Bigg]\Bigg\},
\label{boso_sol2}\hat{\psi}_{R,\uparrow/L,\downarrow}^{({HLL})}(x,t) 
\end{align}
\end{widetext}
where we have identified with $\hat{\psi}^{({ HLL})}_{R,\uparrow/L,\downarrow}(x,t)$ the time-evolution of fermionic operator with respect to the bare $\hat{H}_{HLL}$ without the external gate potential.
Inserting Eqs.~(\ref{bosogate1a})-(\ref{bosogate2a}) into Eq.~(\ref{boso_sol2}) one obtains
\begin{align}
\hat{\psi}_{R,\uparrow/L,\downarrow}(x,t)=\text{exp}\Bigg\{\hspace{1mm}i\frac{eV_0}{2\omega}\Bigg[\alpha_{\sigma}(x)\sin\left(\omega t\right)+\nonumber\\+\beta_{\sigma}(x)\cos\left(\omega
t\right)\Bigg]\Bigg\}\hat{\psi}^{({HLL})}_{R,\uparrow/L,\downarrow}(x,t),\label{psi1}
\end{align}
in the region around the left QPC in Fig.~\ref{giunzione1} at $x\sim-d$, and
\begin{align}
\hat{\psi}_{R,\uparrow/L,\downarrow}(x,t)=\text{exp}\Bigg\{\hspace{1mm}i\frac{eV_0}{\omega}\left(1\pm K\right)\sin\left(\omega\frac{L_2+L_1}{2u}\right))\cdot\nonumber\\\cdot\cos\left[\omega\left(t-\frac{L_2-L_1+2x}{2u}\right)\right]\Bigg\}\hat{\psi}^{({HLL})}_{R,\uparrow/L,\downarrow}(x,t),\label{psi2}
\end{align}
in the region around the right constriction located at $x=d$. In Eq. \eqref{psi1} we defined the geometrical factors
\begin{subequations}
\label{rel_giunzione1}
\begin{align}
\alpha_{\uparrow/\downarrow}(x)&=1-\cos{\left[\omega\left(\frac{L_2+L_1}{2u}\right)\right]}\cos{\left[\omega\left(\frac{x}{u}+\frac{L_2-L_1}{2u}\right)\right]}+\nonumber\\&\pm
K
\sin{\left[\omega\left(\frac{L_2+L_1}{2u}\right)\right]}\sin{\left[\omega\left(\frac{x}{u}+\frac{L_2-L_1}{2u}\right)\right]},\label{alpha}\\
\beta_{\uparrow/\downarrow}(x)&=\sin{\left[\omega\left(\frac{L_2+L_1}{2u}\right)\right]}\cos{\left[\omega\left(\frac{x}{u}+\frac{L_2-L_1}{2u}\right)\right]}+\nonumber\\&\pm
K\cos{\left[\omega\left(\frac{L_2+L_1}{2u}\right)\right]}\sin{\left[\omega\left(\frac{x}{u}+\frac{L_2-L_1}{2u}\right)\right]}.\label{beta}
\end{align}
\end{subequations}
Eqs \eqref{psi1} and \eqref{psi2} can be expressed as
\begin{equation}
\hat{\psi}_{R,\uparrow/L,\downarrow}(x,t)=e^{iA_{2 \sigma}(x)\cos\left[\omega t-\theta_{2 \sigma}(x)\right]}\hat{\psi}^{({HLL})}_{R,\uparrow/L,\downarrow}(x,t),
\end{equation}
in the first region around $x\sim - d$, and
\begin{equation}
\hat{\psi}_{R,\uparrow/L,\downarrow}(x,t)=e^{iA_{1 \sigma}\cos\left[\omega
t-\theta_{1 \sigma}(x)\right]}\hat{\psi}^{({HLL})}_{R,\uparrow/L,\downarrow}(x,t),\label{new_fermionicR22}
\end{equation}
in the second region around $x\sim d$. We have defined the following amplitudes and phase factors:
\begin{align}
A_{2 \sigma}(x)=&\frac{eV_0}{2\omega}\sqrt{\alpha_{\sigma}^2(x)+\beta_{\sigma}^2(x)},\\
\theta_{2 \sigma}(x)=&\arctan{\frac{\alpha_{\sigma}(x)}{\beta_{\sigma}(x)}}+2m\pi
\hspace{5mm}\left(\text{with } m \text{ integer}\right),\\
A_{1 \uparrow/1 \downarrow}=&\frac{eV_0}{\omega}(1\pm K)\sin\left[\frac{\omega}{2u}\left(L_2+L_1\right)\right],\\
\theta_{1 \sigma}(x)=&\frac{\omega}{2u}\left(L_2-L_1+2x\right).
\end{align}
Notably it turns out that the two amplitudes $A_{1 \uparrow}$ and $A_{1 \downarrow}$ do not depend on the position $x$ in the chosen geometry.

\section{Useful integrals}
\label{formule}
In the low temperature limit $k_{{\rm B}} T\ll \omega_{\rm c}$ and $\omega_{\rm c} t\gg1$, Eq. \eqref{boso_correlator} becomes
\begin{equation}
\mathcal{W}(t)=\ln\frac{\pi k_{{\rm B}} T t}{\sinh\left(\pi k_{{\rm B}} T t\right)(1+i\omega_c t)}.
\end{equation}
In this case, it is possible to perform the integral in Eq. \eqref{particle_current} and one obtains
\begin{align}
&\int d\tau \sin\left(n\omega \tau\right) P_{\gamma}\left(\tau-\frac{2d}{u}\right)P_{\gamma}\left(\tau+\frac{2d}{u}\right)=\nonumber\\&\frac{1}{2i} \mathcal{H}\left(\gamma, \frac{2d}{u}n\omega, \frac{2d}{u}k_{\rm B} T\right)\left[\tilde{\mathcal{P}}_{2\gamma}(n\omega)-\tilde{\mathcal{P}}_{2\gamma}(-n\omega)\right],
\end{align}
where
\begin{widetext}
\begin{equation}
\label{AppH}
\mathcal{H}\left(\gamma,x,y\right)=2\pi\frac{\Gamma\left(2\gamma\right)}{\Gamma\left(\gamma\right)}\frac{e^{-2\pi \gamma y}}{\sinh\left(\frac{x}{2y}\right)}\Im\left[\frac{e^{ix}}{\Gamma\left(\gamma+i\frac{x}{2\pi y}\right)\Gamma\left(1-i\frac{x}{2\pi y}\right)}{}_2F_{1}\left(\gamma,\gamma-i\frac{x}{2\pi y},1-i\frac{x}{2\pi y};e^{-4\pi y}\right)\right],
\end{equation}
and
\begin{equation}
\label{AppP}
\tilde{\mathcal{P}}_{g}(\omega)=\left(\frac{2\pi k_B T}{\omega_c}\right)^{g-1}\frac{e^{\frac{\omega}{2k_B T }}}{\omega_c}\mathcal{B}\left[\frac{g}{2}-i\frac{\omega}{2\pi k_B T},\frac{g}{2}+i\frac{\omega}{2\pi k_B T}\right].
\end{equation}
Moreover, by noticing that
\begin{align}
&\int d\tau \sin\left(n\omega \tau\right)\left( \partial_{\tau}P_{\gamma}\left(\tau-\frac{2d}{u}\right)P_{\gamma}\left(\tau+\frac{2d}{u}\right)-P_{\gamma}\left(\tau-\frac{2d}{u}\right)\partial_{\tau}P_{\gamma}\left(\tau+\frac{2d}{u}\right)\right)=\nonumber\\&=-\frac{u}{2}\partial_{d}\int d\tau \sin\left(n\omega \tau\right) P_{\gamma}\left(\tau-\frac{2d}{u}\right)P_{\gamma}\left(\tau+\frac{2d}{u}\right),
\end{align}
the integral in Eq. \eqref{heat_current} can be expressed in the following form
\begin{align}
&\int d\tau \sin\left(n\omega \tau\right)\left( \partial_{\tau}P_{\gamma}\left(\tau-\frac{2d}{u}\right)P_{\gamma}\left(\tau+\frac{2d}{u}\right)-P_{\gamma}\left(\tau-\frac{2d}{u}\right)\partial_{\tau}P_{\gamma}\left(\tau+\frac{2d}{u}\right)\right)\nonumber=\\&=-\frac{u}{4i}\Xi\left(\gamma,2d,\frac{n\omega}{u}, \frac{k_B T}{u}\right)\left[\tilde{P}_{2\gamma}(n\omega)-\tilde{P}_{2\gamma}(-n\omega)\right],
\end{align}
\end{widetext}
where
\begin{equation}
\Xi\left(\gamma,2d,\frac{n\omega}{u}, \frac{k_B T}{u}\right)\equiv\partial_d\left[\mathcal{H}\left(\gamma,\frac{2d}{u}n\omega,\frac{2d}{u}k_B T\right)\right].\label{AppJ}
\end{equation}

\end{document}